# Field-free spin-orbit torque switching enabled by interlayer Dzyaloshinskii–Moriya interaction

Wenqing He[1,2†], Caihua Wan[1,3†*], Cuixiu Zheng[4], Yizhan Wang[1], Xiao Wang[1], Tianyi Ma[1], Yuqiang Wang[1], Chenyang Guo[1], Xuming Luo[1], Maksim. E. Stebliy[5], Guoqiang Yu[1,3], Yaowen Liu[4], Alexey V. Ognev[5], Alexander S. Samardak[5], Xiufeng Han[1-3*]

[1]*Beijing National Laboratory for Condensed Matter Physics, Institute of Physics, University of Chinese Academy of Sciences, Chinese Academy of Sciences, Beijing 100190, China*

[2]*Center of Materials Science and Optoelectronics Engineering, University of Chinese Academy of Sciences, Beijing 100049, China*

[3]*Songshan Lake Materials Laboratory, Dongguan, Guangdong 523808, China*

[4]*School of Physics Science and Engineering, Tongji University, Shanghai, China.*

[5]*Laboratory of Spin-Orbitronics, Institute of High Technologies and Advanced Materials, Far Eastern Federal University, Vladivostok 690922, Russia*

*Corresponding author. xfhan@iphy.ac.cn; wancaihua@iphy.ac.cn

†These two authors contributed equally to this work.

Abstract:

Perpendicularly magnetized structures that are switchable using a spin current under field-free conditions can potentially be applied in spin-orbit torque magnetic random-access memory (SOT-MRAM). Several structures have been developed; however, new structures with a simple stack structure and MRAM compatibility are urgently needed. Herein, a typical structure in a perpendicular spin-transfer torque MRAM, the Pt/Co multilayer and its synthetic antiferromagnetic counterpart with perpendicular magnetic anisotropy, was observed to possess an intrinsic interlayer chiral interaction between neighboring magnetic layers, namely the interlayer Dzyaloshinskii–Moriya interaction (DMI) effect. Furthermore, using a current parallel to the eigenvector of the interlayer DMI, we switched the perpendicular magnetization of both structures without a magnetic field, owing to the additional symmetry-breaking introduced by the interlayer DMI. This SOT switching scheme realized in the Pt/Co multilayer and its synthetic antiferromagnet structure may open a new avenue toward practical perpendicular SOT-MRAM and other SOT devices.

Main text:

Perpendicular magnetization switching driven by spin-orbit torque (SOT) under field-free conditions is crucial for the development of SOT-based magnetic random-access memory (MRAM) [1-6], programmable spin logics [7-9], and neuromorphic computing devices [10, 11] with high density, speed, endurance, and thermal stability. However, owing to the symmetry constrain [12, 13], a magnet with perpendicular magnetic anisotropy (PMA) cannot be deterministically switched by an ordinary spin current with in-plane polarization, unless an extra symmetry breaking factor is introduced. These factors include wedge structures [14, 15], exchange bias/coupling structures [16-20], gradient materials [21, 22], single-crystal materials with low atomic [23, 24] or antiferromagnetic [25, 26] symmetries, and a simple in-plane dipole field [1, 3, 6]. Among them, only the dipolar field [3, 6] and exchange coupling [4, 5] methods have been demonstrated to be integrable with a perpendicular magnetic tunnel junction (MTJ), which is the key unit of SOT devices. Most methods are limited in their compatibility with MRAM technologies; thus, new symmetry-breaking physics and simpler MRAM-compatible structures for hosting field-free SOT-switching are still in demand.

Among PMA materials, magnetic multilayers, such as [Pt/Co]$_n$, are important in memory applications and can be used to develop novel magnetic textures [27, 28]. In such metallic heterostructures, a well-known coupling style is the Ruderman–Kittel–Kasuya–Yosida (RKKY) interaction [29], which enables the ferromagnetic or antiferromagnetic coupling of adjacent magnetic layers through nonmagnetic spacers. For example, the synthetic antiferromagnet (SAF) of [Pt/Co]$_n$/Ru/[Co/Pt]$_n$ is not used directly as the ferromagnetic electrodes of MTJs; however, it has been widely utilized as a pinned layer in spin-transfer torque MRAM, owing to its negligible dipole field and high thermal stability [30, 31]. It is worth noting that perpendicular SAF structures can be switched by SOT with the assistance of an external magnetic field [30, 31]; however, apart from a wedge structure [15], this has rarely been reported in field-free conditions.

In addition to the RKKY interaction, another magnetic coupling style, the Dzyaloshinskii−Moriya interaction (DMI), recently gained research interest. It is an antisymmetric exchange interaction that favors the vertical coupling of neighboring spins with spatial inversion asymmetry [32-43]. This stems from an additional RKKY term, owing to the three-site spin-orbit scattering of conduction electrons by non-magnetic atoms, as elaborated by Fert and Levy [32]. In addition to bulk PMA materials [33], DMI has also been predicted and confirmed in interface systems with broken inversion symmetry [34-37]. This interfacial DMI helps to stabilize chiral spin textures, such as skyrmions [36, 37] or chiral domain walls [38-40]. It also eases domain-wall nucleation at the boundary in the switching dynamics of nanomagnets driven by SOT [41-43] and facilitates field-free spin logic in laterally engineered nanomagnets [9, 40].

Recently, atomistic Monte Carlo calculations showed that the three-site scattering model can be generalized to ferromagnet (FM)/non-magnet (NM)/ferromagnet (FM) trilayers, in which the DMI mediates chiral coupling between the FM layers [44]. Spin directions of the two FM layers are denoted by $\mathbf{S}_T$ and $\mathbf{S}_B$ hereafter (Figure 1). Their coupling harvests an energy $E_{DM}$ via $E_{DM}=$

-**D**·(**S**$_T$×**S**$_B$). Vector **D** characterizes the magnitude of the interlayer DMI and determines the plane normal unfolded by **S**$_T$ and **S**$_B$. This novel effect was experimentally observed in (canted) SAF structures [45-48] and a magnetic trilayer with an in-plane anisotropic layer, a PMA layer, and a coupling spacer [47], which led to nontrivial magnetization behaviors. For example, Avci *et al*. [47] reported the occurrence of a strong interlayer DMI in the order of 160 Oe between an in-plane magnetized Co layer and a perpendicularly magnetized TbFe layer through a Pt spacer. The perpendicular (in-plane) TbFe (Co) layer could be magnetized by an in-plane (out-of-plane) field, owing to the interlayer DMI. Such a strong DMI coupling can be harnessed to design vertical heterostructures with correlated magnetizations for logic applications [47]. These reports reveal that the recently discovered interlayer DMI can potentially enable novel 3-dimensional spin textures, thereby creating and manipulating unprecedented magnetic effects in PMA multilayers [45].

SOT-driven spin dynamics combined with interfacial DMI [9, 40, 49] as well as field-driven magnetizing behaviors manipulated by interlayer DMI [45-48] have been previously studied; however, the concerted interplay of the interlayer DMI with SOT, which can result in nontrivial spin dynamics driven by electric methods, has not yet been investigated. Herein, we identify that the interlayer DMI can effectively reduce system symmetry and aid SOT to enable (prohibit) field-free switching of perpendicular magnetization when current **j** is applied colinear (vertical) to the **D** vector. This switching can be interpreted using a macrospin model and micromagnetic simulations. Field-free and interlayer-DMI-enabled magnetization switching achieved in MRAM-friendly structures may advance the development of practical SOT devices.

First, we analyze the symmetry of a dual-spin system with interlayer DMI. The emergence of the interlayer DMI inevitably reduces the system symmetry (Figures 1a-b). Owing to its pseudo-vector nature, a spin vector retains (reverses) its direction as mirrored by a plane that is perpendicular (parallel) to the spin [14]. Therefore, upward and downward spins are symmetric for any Z-parallel mirrors. However, for an interlayer DMI system with an in-plane **D** vector [45-48], its symmetry gets reduced. There are only two candidate mirrors remaining, one (Mir$_{//}$) parallel and the other (Mir$_\perp$) vertical to the **D** vector (Figure 1a-b). We first analyze Mir$_{//}$. **S**$_T$ and **S**$_B$ in the left panel of Figure 1a become **S**$_T$' and **S**$_B$' on the right when the system is mirrored about Mir$_{//}$. For the original and mirrored systems, the interlayer DMI energies are opposite. $E_{DM}$<0 for the original, but $E_{DM}'=-E_{DM}$>0 for the mirrored system, which causes Mir$_{//}$ symmetry breaking by the interlayer DMI. However, **S**$_T$ and **S**$_B$ in the left panel of Figure 1b become **S**$_T$' and **S**$_B$' on the right after the Mir$_\perp$ operation. In this case, $E_{DM}'=E_{DM}$. The Mir$_\perp$ mirror symmetry was retained. Thus, an extra factor is required to break the Mir$_\perp$ mirror and determine the final spin-up or spin-down state. A feasible method is to use a current vertical to Mir$_\perp$ (or parallel to **D**). In the following, this symmetry-breaking law (**j**//**D**) to realize field-free switching is verified.

Two series of stacks, Pt(2)/[Co(0.4)/Pt(0.7)]$_n$/Pt(1.3) with $n$=1–6 and Pt(2)/Co(0.5)/Pt(0.7)/Ir($t_{Ir}$)/Pt(0.7)/Co(0.5)/Pt(2), were studied to test the interlayer DMI and the SOT-switching scheme. The numbers in brackets are the thicknesses in nanometers. While the first stack only exhibited ferromagnetic coupling between the Co layers, the other displayed ferromagnetic or antiferromagnetic coupling when the appropriate spacer thickness was chosen. A thickness of $t_{Ir}$=1.3 nm led to the SAF coupling. Except where mentioned, all the samples were grown under an in-plane field, which was the key to controlling the **D** direction of the interlayer DMI and possibly lower in-plane symmetry.

We present data from S1: Pt/Co/Pt/Co/Pt and S2: Pt/Co/Pt/Ir(1.3)/Pt/Co/Pt as corresponding examples for the ferromagnetic and SAF structures. S3: Pt(2)/Co(0.4)/Pt(2) and S4: Pt(2)/Co(0.4)/Pt(0.7) stacks deposited in the same chamber as S1/S2 were used as control samples. All samples had out-of-plane easy axes, owing to the high interfacial PMA of the Pt/Co and Co/Pt interfaces. The saturated magnetization $M_T$ and $M_B$ for S1 and S2 were obtained from the *M-H* loops (Supplement I). After the stacks were patterned into crossbars, PMA was further confirmed using $R_{xy}$-$H_z$ loops (Figure 2), where $R_{xy}$ represents the anomalous Hall resistance. From the $R_{xy}$-$H_{z/x}$ loops, the anisotropic field and exchange-coupling energy were retrieved for the following micromagnetic simulation (Supplement I).

Taking S1 as an example, we characterized the interlayer DMI using the loop-shift method [46] and determined its **D** vector. For simplicity, we reformed $E_{DM}$ as $E_{DM} \equiv -\mathbf{D} \cdot (\mathbf{S}_T \times \mathbf{S}_B) = -\mathbf{S}_T \cdot (\mathbf{S}_B \times \mathbf{D})$. An effective field $\mu_0\mathbf{H}_{DM,T}=(\mathbf{S}_B \times \mathbf{D})/(M_T t_T)$ acted on $\mathbf{S}_T$, owing to the interlayer DMI. $t$ ($\mu_0$) is the thickness (permeability of vacuum). If $E_{DM} \ll K_B$ (effective PMA energy), $\mathbf{S}_B$ was approximately pointed in the ±z direction with a negligible tilt, or $S_B \approx \pm e_z$ where $e_z$ was the corresponding unit vector. If a small in-plane field **H** was applied, $S_B$ was then tilted by a small angle and $S_B \approx \pm e_z + H/H_{KB}$, where $H_{KB} \equiv 2K_B/M_B$. Then, $(\mu_0 M_T t_T)H_{DM,T} \approx \pm e_z \times D + (H \times D)/H_{KB}$. Because **D** lies in-plane [45-48], the first term in $H_{DM,T}$ was also in-plane. In contrast, the second term contributed a z-component to $H_{DM,T}$, which was maximized (minimized) as $H \perp D$ ($H // D$). Similar arguments hold for the other layer. Owing to this z-component, the $R_{xy}$-$H_z$ loops were offset by the in-plane **H**. We used $H_o = (H_{sw}^{DU} + H_{sw}^{UD})/2$ to characterize the offset with $H_{sw}^{DU}$ ($H_{sw}^{UD}$) of the switching field from the spin-down to spin-up (spin-up to spin-down) state. Noticeably, the maximized $\mu_0 H_o = HD/(t_T M_T H_{KB})$ as $H \perp D$.

In the experiment, we applied a prefixed in-plane $H_{in}$ and measured $R_{xy}$ as a function of $H_z$ (Figure 2a). After acquiring each $R_{xy}$-$H_z$ loop, $H_{in}$ was rotated at an azimuthal angle $\varphi$. During the rotation process, the offset was able to be maximized (Figure 2b-c). Figure 2c shows $|H_{sw}|$ as a function of $\varphi$. $H_o$ was maximized in a particular direction (asymmetric axis) and minimized in the vertical direction (symmetric axis). Here, the symmetric (asymmetric) axis was defined by the line along which $|H_{sw}|$ differs the least (the largest). Correspondingly, **D** was then determined along the symmetric axis. The maximum $H_o$=8.3 Oe (14.3 Oe) at $H_{in}$=30 Oe (50 Oe). According to the formula $(H_o/H_{in})=D/(\mu_0 t_T M_T H_{KB})=0.28$, $E_{DM}=|D|=\mu_0(H_o/H_{in})(H_{KB}M_T t_T)=11.3$ μJ/m² for S1, which is of the same order as the result in Ref. [47]. Notably, the direction of **D** was

perpendicular to the magnetic field applied during the deposition process, which was necessary to control the interlayer DMI effect, as discussed later in this Letter.

This interlayer DMI was also observed for S2 with a SAF structure. Following the same procedure, we obtained two polar $|H_{sw}|$ diagrams for the two switching events in S2. Figures 2e-f show the same symmetric and asymmetric axes, thereby indicating the **D** direction for S2. For the top and bottom layers of S2, the maximum offset $H_o$ was approximately 11.2 and 13 Oe, respectively, under the condition of $H_{in}$=100 Oe. $E_{DM,j}$ was estimated using $\mu_0(H_{oj}/H_{in})(H_{Ki}M_jt_j)$, as described above. Thus, $E_{DM,B}$=5.1 µJ/m² and $E_{DM,T}$=4.9 µJ/m², closely equal for both layers as expected. Owing to the thicker spacer (1.4 nm Pt and 1.3 nm Ir), reasonably, S2 had a smaller interlayer DMI energy than S1. We also performed the same measurements for Pt(2)/Co(0.4)/Pt(2 nm). No offset was observed, which indicated its relevance to the interlayer coupling (Supplement II).

Notably, a magnetic field of 180 Oe was applied during the film deposition, and its direction was perpendicular to the subsequently determined **D** direction. The method of using an induced field to engender the interlayer DMI effect was also mentioned in previous studies [46, 47]. Here, we used another sputtering system without an induced field to grow the same stacks in which the interlayer DMI was absent; however, PMA was retained (Supplement III). This contrast indicates that the interlayer DMI was caused by the induced field during deposition. The mechanism of controlling the interlayer DMI using the induced field requires further exploration; however, this technique enables the manipulation of the interlayer DMI to achieve current-induced spin-orbit torque switching in a controlled manner.

Then, we attempted to switch the perpendicular magnetization of the Pt/Co multilayers and their SAF counterparts by SOT following the symmetry-breaking law of **j**//**D**. To verify the physical scenario shown in Figure 1, it is better to realize parallel and vertical configurations between the applied current density (**j**) and **D**. We could apply $I_x=I_0\cos\varphi_j$ and $I_y=I_0\sin\varphi_j$ simultaneously, owing to the crossbar geometry (Figure 3a). Then, the **j** direction ($\varphi_j$ angle) in the cross-region could be controlled by $I_x$ and $I_y$. $\varphi_j=0$ corresponded to the predetermined **D** direction. By applying **j**, two spin currents with opposite spin polarizations $\sigma_{T/B}$ were generated in the upper and lower platinum layers via the spin Hall effect and then transferred to the upper and lower cobalt layers, respectively. Finally, the damping-like spin torques of $\mathbf{S}_{T/B}\times\sigma_{T/B}\times\mathbf{S}_{T/B}$ were imposed on both layers, and their spin dynamics were activated. Here the damping-like and field-like torque efficiencies were measured (Supplement IV) and the former dominated in our Pt/Co system, similar with Ref. [50].

We first consider the S1 sample. SOT switching loops for various $\varphi_j$ values are shown in Figure 3b. No magnetic field was applied here. In particular, field-free SOT switching was achieved when $\varphi_j \neq 90º$ or $270º$. We summarize the switching degree ($\Delta R$ at $I_0=0$ mA) and the critical current ($J_c$) as functions of $\varphi_j$ in Figure 3c. These diagrams confirm that $\Delta R_{xy}$ ($J_c$) was maximized (minimized) as **j**//**D**, and $\Delta R_{xy}$ ($J_c$) became zero (divergent) as **j** $\perp$ **D**. The maximum switching degree reached 51.2%, which may be due to shunting of the crossbar structures. We also attempted to verify the switching law (**j**//**D**) by patterning crossbars with different orientations. The results also supported the observed trend though their switching degree differed a little (Supplement V). Field-free SOT switching was also performed for S2. As **j**//**D**, switching loops with odd symmetry between two SAF states were observed (Figure 3d), and the switching degree reached 50.8%; however, the switching was absent as **j**⊥**D** (Figure 3e). The same trend was observed for the SAF sample. The SOT-switching performance under an external field for S1 and S2 are also attached in Supplement VI for comparison.

To further confirm the symmetry-breaking law of **j**//**D** and exclude other symmetry breaking factors, we also tested S3: Pt(2)/Co(0.4)/Pt(2) and S4: Pt(2)/Co(0.4)/Pt(0.7) control samples without interlayer DMI for comparison (Supplement II). S3/S4 were deposited in the same sputtering chamber and following the same sputtering conditions with S1 and S2. For the symmetric S3, its perpendicular magnetization could not be switched because the spin currents from both Pt layers had opposite polarizations and canceled each other. However, for the asymmetric S4, SOT-driven magnetization switching was achieved using a **j**-parallel field, which is classified as a Type-Z scheme [13] (perpendicular easy axis and current-parallel field). However, field-free switching was not observed for S3 and S4, thereby confirming the relevance of the field-free switching with the interlayer coupling.

The SOT switching in the interlayer DMI system can also be understood as follows (Figure 4a). $E_{DM}=-\mathbf{D}\cdot(\mathbf{S}_T\times\mathbf{S}_B)$ can be rewritten as $E_{DM}=-2\mathbf{D}\cdot(\mathbf{s}\times\mathbf{S})=-2\mathbf{S}\cdot(\mathbf{D}\times\mathbf{s})$, with $\mathbf{s}\equiv(\mathbf{S}_T-\mathbf{S}_B)/2$ and $\mathbf{S}\equiv(\mathbf{S}_T+\mathbf{S}_B)/2$. Clearly, $\mathbf{s}$ is vertically and chirally coupled with $\mathbf{S}$ via the interlayer DMI. After applying spin currents ($\boldsymbol{\sigma}_T$ and $\boldsymbol{\sigma}_B$) produced in the upper and lower Pt layers by the spin Hall effect, $\mathbf{S}_T$ and $\mathbf{S}_B$ are pulled toward $\boldsymbol{\sigma}_T$ and $\boldsymbol{\sigma}_B$, respectively. When the torques become large enough, $\mathbf{S}_T$ and $\mathbf{S}_B$ are even aligned in-plane, resulting in a non-zero $\mathbf{s}$ that is parallel to $\boldsymbol{\sigma}_T-\boldsymbol{\sigma}_B$ (Figure 4a and Supplement VII). Here, $\boldsymbol{\sigma}$ denotes the polarization direction of spin currents. As $\mathbf{j}//\mathbf{D}$, $\mathbf{s}$ becomes vertical to $\mathbf{D}$. Thus, $E_{DM}$ is reduced by choosing the preferred $\mathbf{S}$ direction as $\mathbf{S}//(\mathbf{D}\times\mathbf{s})$. An opposite $\mathbf{j}$ resulted in an opposite $\mathbf{s}$ and finally opposite $\mathbf{S}$, and deterministic switching occurred (Supplement VII). The macrospin model qualitatively reproduced the field-free switching performance and visualized the spin dynamics (Figure 4b).

Interlayer-DMI-enabled SOT switching was also supported by micromagnetic simulations performed using MuMax[3, 51] (Figure 4c-h) (for more details see the Method and Supplement VII). For simplicity, we considered two samples with a Co/Pt/Co trilayer structure. The lateral dimensions of the samples were $400 \times 400$ nm$^2$. In addition to the FM or AFM coupling between the Co layers, interlayer DMI can also be included. A detailed description of the parameters is given in the Method and Supplement VII sections, and most of these parameters were obtained from our experiments. For the FM coupling sample (S1) in Figure 4c, when the current pulse (along the y-axis) is applied between $t=10–30$ ns, the initial parallel state (along the +z-axis) of the two Co layers is quickly aligned in-plane (along the x-axis) by the SOT. After the current pulse, the in-plane magnetization either stochastically relaxes to the +z or -z state for the case without interlayer DMI or switches to the –z-direction for the case with interlayer DMI at 49 ns.

We performed similar simulations for the AFM sample (S2). The initial antiparallel state was a tail-to-tail configuration along the z-axis. After the current pulse, the two Co layers were switched to the antiparallel head-to-head state for the case with interlayer DMI (Figure 4h); however, the layers were randomly switched to the tail-to-tail or head-to-head state for the case without interlayer DMI. Some typical spin dynamics for the entire process are shown in the Supplemental movies. These results agree well with the experimental results, which demonstrates the key role of the interlayer DMI in field-free SOT switching. It should be noted that the intralayer (or interfacial) DMI for Pt(2)/Co(0.4)/Pt(0.7 nm) was small and its presence cannot result in the deterministic switching behavior for S1/S2 in the micromagnetic simulations (Supplement VII).

In addition to S1/S2, similar switching behaviors enabled by the interlayer-DMI were also observed for Pt/[Co/Pt]$_n$/Pt with $n$=3–5. For $n$=2 or 3, field-free switching was still observed at 360 K and 400 K, respectively, which supports their use at ambient temperatures. Furthermore, the SOT switching performance of the $n$=2 sample was checked $10^4$ times without notable degradation. All these characteristics demonstrate the application prospects of this Pt/Co multilayer system with the interlayer DMI (Supplement VIII).

In conclusion, we implemented a field-free SOT-switch of perpendicular magnetization of both ferromagnetic Pt/Co/Pt/Co/Pt and synthetic antiferromagnetic Pt/Co/Pt/Ir/Pt/Co/Pt structures with the interlayer-DMI effect. The symmetry-breaking law of **j**//**D** was experimentally confirmed and theoretically elaborated upon. In comparison to previous field-free switching methods with redundant functional layers or additional asymmetric designs, the interlayer-DMI effect that enables field-free switching is an intrinsic property of Pt/Co systems, which can be controlled by an induced field during deposition. Recently, single-crystal materials with low atomic or magnetic symmetry, such as CuPt [24] and IrMn [25], have been reported to favor field-free switching as spin current sources. Here, Pt/Co multilayers and their SAF equivalents were fabricated using magnetron sputtering technologies on thermally oxidized silicon substrates. They were polycrystalline in nature, and had loose requirements for their crystal structures. Pt/Co multilayers have been applied in the MRAM industry as functional pinned layers and are compatible with prior art. Therefore, the field-free SOT switching realized in these Pt/Co structures with interlayer DMI may offer an opportunity to develop practical SOT devices based on state-of-the-art technologies.

**Materials and Methods**

**Stacks and measurement**: Pt(2)/Co(0.4)/Pt(0.7)/Co(0.4)/Pt(2), Pt(2)/Co(0.5)/Pt(0.7)/Ir(1.3)/ Pt(0.7)/Co(0.5)/Pt(2), and other control sample stacks were deposited using a magnetron sputtering system (ULVAC Inc., Japan) with a base pressure of $3.0 \times 10^{-6}$ Pa and an in-plane field of 180 Oe. The numbers in brackets are the thicknesses in nanometers. Another magnetron sputtering system (AJA International Inc., USA) without an induction field was used to deposit the corresponding control samples. A vibrating sample magnetometer (MicroSense) was used to measure the hysteresis loops. We then patterned the stacks via ultraviolet lithography and subsequent argon-ion etching techniques into crossbar devices with a width and length of 10 μm and 100 μm, respectively. Pt(10)/Au(80) pads were used to connect the crossbar terminals. Two Keithley 2400 and a Keithley 2182 source meters were used to provide source currents and measure the Hall voltage, respectively. The hysteresis loops of the anomalous Hall resistance of the multilayers were measured using an electromagnetic probe workstation equipped with three-dimensional (3D) Helmholtz coils, which enabled the measurement of the anomalous Hall signal on a crossbar while sweeping a magnetic field along the out-of-plane direction and presetting an in-plane magnetic field. A physical-property measurement system (PPMS-9T, Quantum Design) and 3D magnetic field probe station (East Changing Technologies, China) were used to provide magnetic fields during the magnetotransport and SOT-switching experiments. Moreover, to test field-free switching, we placed the devices completely outside the magnetic-field generators to avoid any remanence of the PPMS or 3D Helmholtz coils. We simultaneously applied pulse currents $I_x$ and $I_y$ at two mutually perpendicular terminals to crossbar devices, and the magnitude and direction of the currents in the two directions satisfied the relationships $I_x=I_0\cos\varphi_j$ and $I_y=I_0\sin\varphi_j$. These relations ensured that the direction of the current density could be changed relative to the **D** vector without changing the devices. Except when otherwise mentioned, all measurements were conducted at room temperature.

**Micromagnetic simulations.** Micromagnetic simulations were performed using the GPU-accelerated simulation package Mumax[3, 51], including exchange, PMA, dipolar, and interlayer exchange coupling (IEC) through the RKKY effect, intralayer and interlayer DMI interactions. The code is based on the Landau–Lifshitz–Gilbert equations, which can calculate space- and time-dependent magnetization dynamics. In this study, we considered two samples with a Co (lower)/Pt/Co(upper) trilayer structure: one with FM coupling (S1 sample) and the other with AFM coupling (S2 sample) between the lower (L) and upper (U) Co layers. The lateral dimensions of the sample were $400 \times 400$ nm$^2$, and a discretization cell of $4\times4\times1$ nm$^3$ was used. The typical parameters used for the samples were as follows: $A_{ex} = 13$ pJ/m (exchange stiffness) and $\alpha = 0.05$ (Gilbert damping). For sample S1: saturation magnetization $M_{sU} = M_{sL} = 1.06 \times 10^6$ A/m, perpendicular magnetic anisotropy $K_U = K_L = 3.5\times10^4$ J/m$^3$ (taken from the experimental data), the IEC strength between the two Co layers $A=2.8\times10^{-6}$ J/m$^2$, and the interlayer DMI strength $D=1.13\times10^{-5}$ J/m$^2$. For sample S2: $M_{sU} =1.12\times 10^6$ A/m, $M_{sL}= 1.423 \times 10^6$ A/m, $K_U =3.1\times10^4$ J/m$^3$, $K_L =5.6\times10^4$ J/m$^3$, $A =-2.8\times10^{-6}$ J/m$^2$, and $D=7.5\times10^{-6}$ J/m$^2$. For comparison, the same system but with an additional intralayer DMI of $2.4\times10^{-6}$ J/m$^2$ was also simulated. All simulations were performed without considering the temperature.

ASSOCIATED CONTENT

**Supporting Information:** Supporting information is provided as a separate file.

1. Magnetic structures for the ferromagnetic S1 and SAF S2 samples; 2. Loop-shift and SOT-switching measurement for the control samples; 3. No interlayer DMI effect for the sample deposited without an induced field; 4. Measurement of SOT efficiency for S1 and S4; 5. Two different methods to confirm the field-free switching law of **j**//**D**; 6. Switching performance of the S1/S2 samples under an applied field; 7. Micromagnetic and Macrospin simulations of field-free SOT-switching; 8. Switching performance at higher temperatures and longer endurance and more periods.

**Acknowledgements**

**Funding:** The authors appreciate the financial support from the National Key Research and Development Program of China (MOST) (Grant No. 2017YFA0206200), the National Natural Science Foundation of China (NSFC) [Grant No. 51831012, 11974398, 12061131012], Beijing Natural Science Foundation (Grant No. Z201100004220006), the Strategic Priority Research Program (B) and the Key Research Program of Frontier Sciences and the K. C. Wong Education Foundation of Chinese Academy of Sciences (CAS) [Grant No. XDB33000000 and QYZDJ-SSW-SLH016 and GJTD-2019-14] and Foshan Science and Technology Innovation Team Project [FS0AA-KJ919-4402-0022]. C. H. Wan appreciates financial support from Youth Innovation Promotion Association, CAS (2020008). The work of M.E.S. and A.S.S. related to the modelling and theoretical analysis was supported by the Russian Science Foundation (Project No. 21-42-00041). A.V.O. thanks the Russian Ministry of Science and Higher Education for state support of scientific research conducted under the supervision of leading scientists in Russian institutions of higher education, scientific foundations and state research centers (Project No. 075-15-2021-607).

**Author contributions:** X.F.H. led and was involved in all aspects of the project. W.Q.H, C.H.W., C.Y.G, T.Y.M, and X.M.L. deposited stacks and fabricated devices. W.Q.H, C.H.W, X.W, Y.Z.W, Y.Q.W conducted magnetic and transport property measurement. C.H.W., W.Q.H., C. X. Z, Y. W. L., M. E. S. contributed to modelling and theoretical analysis. C.H.W., W.Q.H., G.Q.Y., A. S. S., A. V. O., Y. W. L., and X.F.H. wrote the paper. X.F.H. and C.H.W. supervised and designed the experiments. All the authors contributed to data mining and analysis.

**Competing interests:** The authors declare no competing interests.

**Data availability:** Data supporting the findings of this study are available from the corresponding authors on a reasonable request.

(51) Vansteenkiste, A.; Leliaert, J.; Dvornik, M.; Helsen, M.; Garcia-Sanchez, F.; Van Waeyenberge, B. The design and verification of MuMax3. *AIP Advances* **2014**, *4* (10). DOI: 10.1063/1.4899186.

**Figures:**

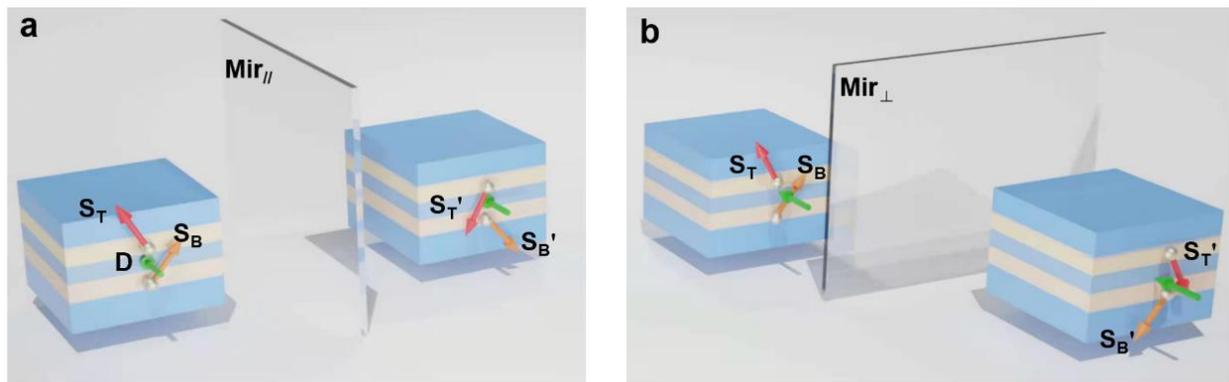

**Figure 1.** Symmetry breaking of an interlayer DMI system. (a) Symmetry breaking by the **D**-parallel mirror Mir$_\parallel$. The original (left) and mirrored (right) states have opposite interlayer DMI energy. (b) Symmetry maintaining using the **D**-vertical mirror Mir$_\perp$. Notably, for the same sample, the **D** vector of the interlayer DMI effect should be the same, as shown by the green arrows.

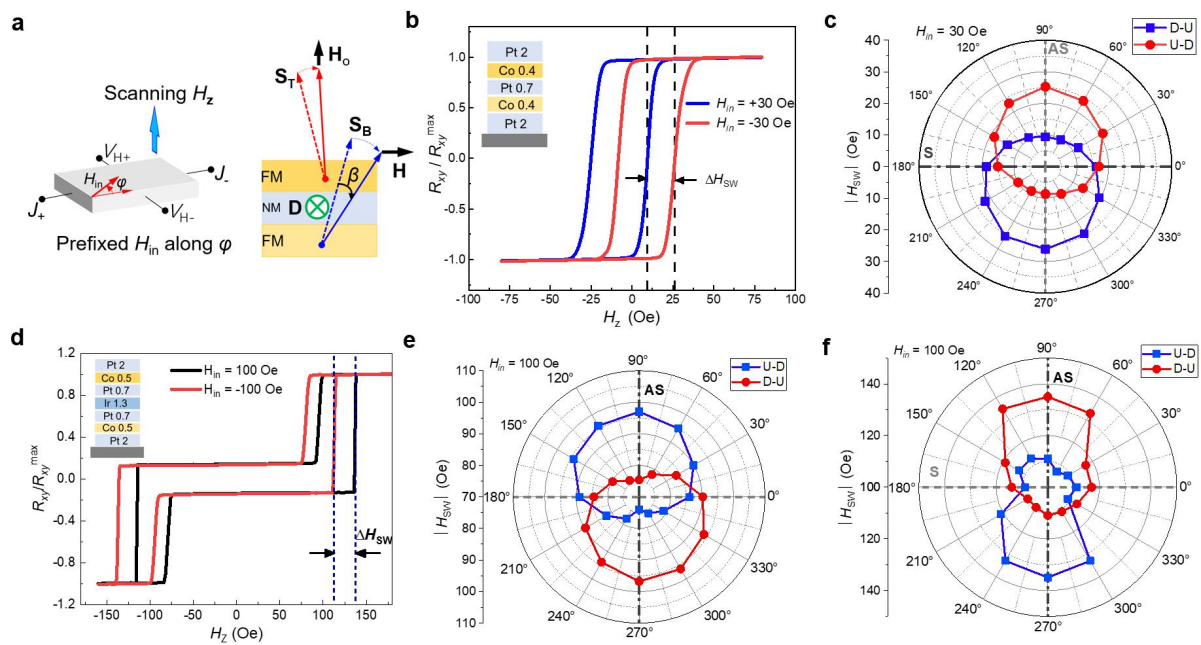

**Figure 2.** Characterization of the interlayer DMI effect for the ferromagnetic and SAF structures. (a) Experimental setup for the loop-shift measurement and a diagram depicting how an in-plane $H$ generates an out-of-plane offset $H_o$ by the interlayer DMI. (b) and (d) Maximum loop shifts for the ferromagnetic and SAF samples, respectively. (c) and (e-f) Polar diagrams of the switching field $H_{sw}$ on the $\varphi$ angle of the in-plane field for the ferromagnetic and SAF samples, respectively. (e-f) correspond to the two switching events for the SAF sample. These polar diagrams show asymmetric (AS) and symmetric (S) axes. The **D** vector of the interlayer DMI effect can be distinguished as parallel to the S axis.

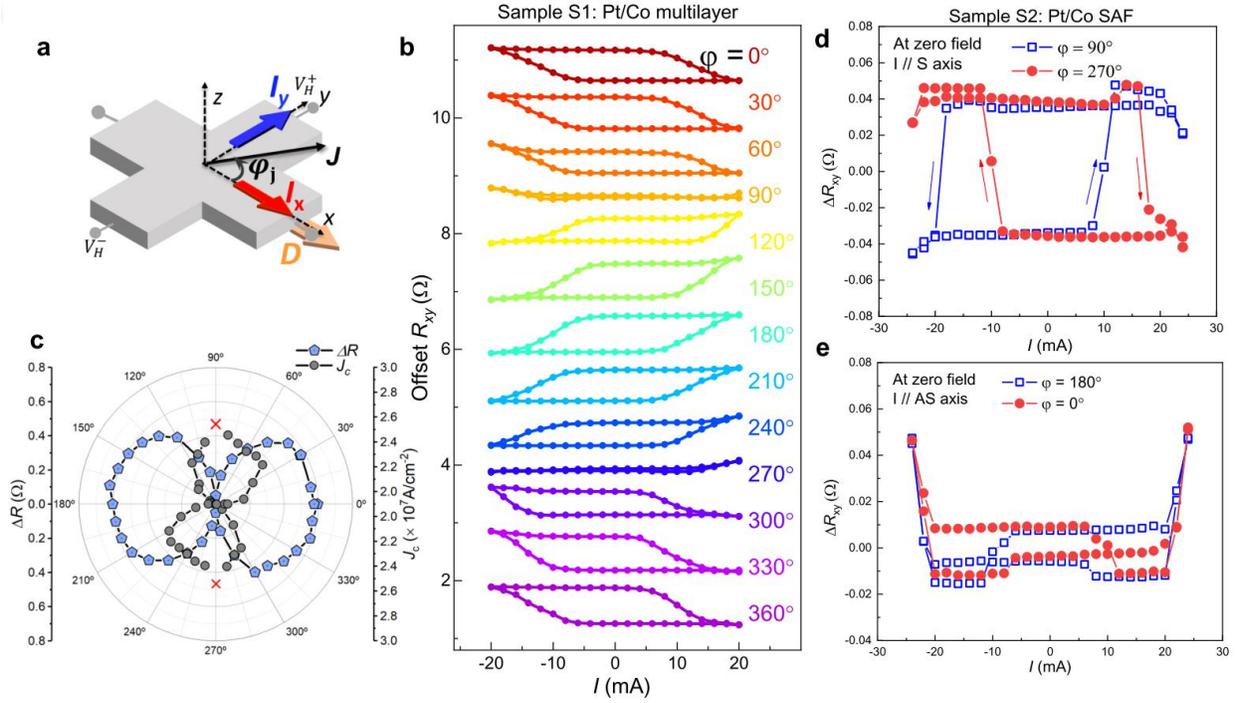

**Figure 3.** Field-free SOT switching enabled by the interlayer DMI. (a) SOT-switching measurement diagram. $I_x$ and $I_y$ were applied simultaneously to adjust the direction ($\varphi_j$) of the current density in the cross region relative to the **D** vector. $\varphi_j=0$ marked the direction parallel to **D**. (b) Field-free SOT switching at different $\varphi_j$ for the ferromagnetic sample S1. The switching loops disappeared at $\varphi_j=90°$ and $270°$. (c) Resistance change ($\Delta R$) and critical current ($J_c$) of the switching as functions of $\varphi_j$ for S1. (d-e) Field-free SOT-switching for the SAF S2 sample with **j**//**D** and **j**⊥**D**, respectively.

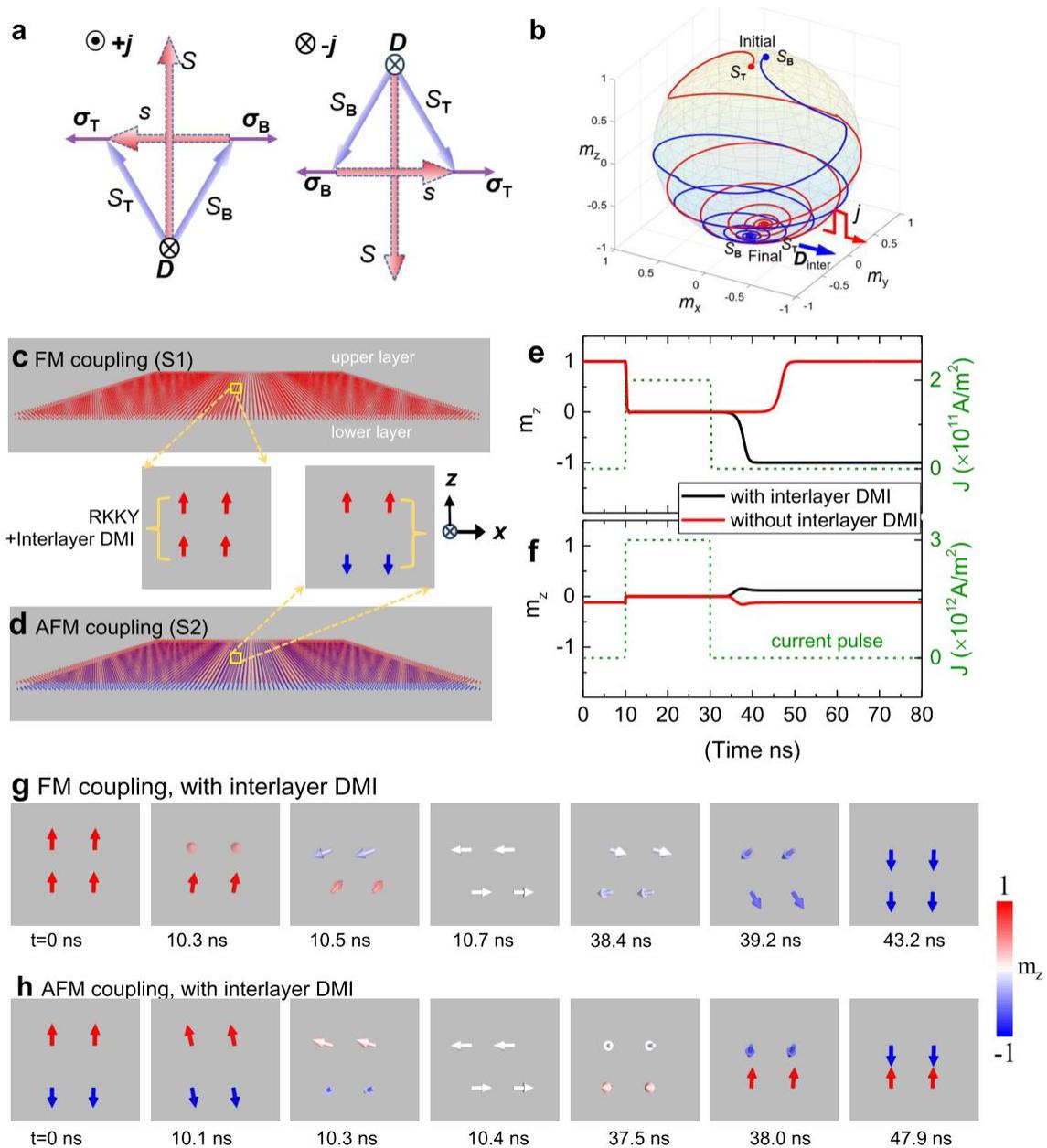

**Figure 4.** Switching mechanism and simulation of field-free SOT switching enabled by interlayer DMI. (a) Diagrams to depict the chiral coupling between **s** and **S** and the relationship between **s** and **j**. **j** determines the spin polarizations $\sigma_{T/B}$ and thus, the direction of **s**. The direction of **S** can be controlled via the chiral coupling with **s**. (b) Switching dynamics activated

by a pulse current parallel to **D**. (c-h) Micromagnetic simulations of the field-free SOT-switching. (c-d) Simulation models comprising two Co layers with FM or AFM coupling with the interlayer DMI. (e-f) Current-pulse (green dashed curves)-induced time evolution of magnetization precession ($m_z$ component) for the two samples; the black and red curves indicate the cases with and without the interlayer DMI, respectively. (g-h) Transient snapshots of magnetization during the switching process for the two samples with the interlayer DMI. Here, we show an enlarged view with four spins taken from the samples. The color of the arrows indicates the direction of the z-component of the magnetization. The snapshots are from the x-z plane, whereas the current pulse is applied along the y-axis. Figure 4g matches the picture in Figure 4a.